\documentclass[epj]{svjour}
\usepackage{graphics}
\begin{document}
\title{Conserved Dynamics and Interface Roughening in Spontaneous
Imbibition : A Critical Overview
}
\author{M. Dub\'e \inst{1,2} \thanks{\emph{Present address:
Center for the Physics of
Materials, McGill University, 3600 rue University, Montr\a'eal, Qu\a'ebec,
Canada H3A 2T8   }   }, M. Rost \inst{1,2} \and M. Alava\inst{2}
}                     
%
%
\institute{Helsinki Institute of Physics, P.O. Box 9
(Siltavuorenpenger 20 C), FIN--00014 University of Helsinki,
Helsinki, Finland \and Laboratory of Physics, Helsinki University 
of Technology, P.O. Box 1100, FIN--02150 HUT, Espoo, Finland
}
\authorrunning{M. Dub\a'e, M. Rost and M. Alava}
\titlerunning{Conserved Dynamics in Imbibition}
%
%
\abstract{
Imbibition phenomena have been widely used experimentally and
theoretically to study the kinetic roughening of interfaces. We
critically discuss the existing experiments and some associated
theoretical approaches on the scaling properties of the imbibition
front, with particular attention to the conservation law associated to
the fluid, to problems arising from the actual structure of the
embedding medium, and to external influences such as evaporation and
gravity. Our main conclusion is that the scaling of moving interfaces 
includes many crossover phenomena, with competition between the average 
capillary pressure gradient and its fluctuations setting the maximal 
lengthscale for roughening. We discuss the physics of both pinned and
moving interfaces and the ability of the existing models to account for 
their properties.
} 
\maketitle
\section{Introduction}
\label{intro}

A considerable amount of effort has been spent in the field of kinetic
roughening over the last two decades. Apart from technological
interests in crystal growth, a good part of the fascination for this
field comes from the possibility of describing many different types of
interfaces by a few distinct universality classes, in terms of scaling
exponents and scaling functions \cite{ref_gen,Krug_97}. 

However, if the theoretical and numerical aspects of the field are
extremely rich and varied, the experimental backing of these ideas is
quite lacking. This is for instance the case in kinetic roughening in
crystal growth due to the great variety of atomistic processes. Much
of the experimental attention has rather concentrated on simpler 
one--dimensional ``toy-systems'', in particular driven interfaces in random
media. A definite advantage of these systems is that the interface
configuration is directly observable in the experiment (while, it must
often be deduced, from some probe-surface interaction, in crystal
growth). Even these simple systems are challenging. They are most 
commonly grouped in universality classes described by the Kardar-Parisi-Zhang
\cite{KPZ_86} (KPZ) or Edwards-Wilkinson \cite{EW_82} (EW) equations,
with either quenched or thermal noise, depending on 
the driving regime. In the case of quenched noise the exact 
low-dimensional scaling behaviour is however still somewhat contradictory
\cite{Leschhorn_96,Nattermann_94,Leschhorn_97,Narayan_93,Csahok_93}).

Examples of experimental studies in these systems are slow combustion
fronts \cite{jyvkl_97} (for which KPZ scaling was recently demonstrated), 
shock fronts, fluid-gas interfaces in Hele-Shaw cells
\cite{Rubio_89,Horvath_91a,He_92b,Delker_96,Dougherty_98}, or in paper
\cite{Buldyrev_92,Barabasi_92,Family_92,Buldyrev_ph,Amaral_94,Horvath_95,Kumar_96,Kwon_96,Zik_97}
and fracture surfaces \cite{Bourev} such as one-dimensional fracture
lines \cite{Kertesz_93,Engoy_94}, to name a few. In this paper, we
concentrate on a particular system, namely the spontaneous imbibition
of a porous medium by a liquid
\cite{Buldyrev_92,Barabasi_92,Family_92,Buldyrev_ph,Amaral_94,Horvath_95,Kumar_96,Kwon_96,Zik_97}.
Our goal is to critically review the experiments and theories that
exist in the literature and to indicate a direction for further
investigation and comprehensive understanding of imbibition front
roughening. The key questions are (i) whether imbibition should
present any kind of scaling behaviour at all, and (ii), if so, under
which conditions can one of the -- possibly several -- scaling regimes
become observable. 

Our motivations in writing this paper are twofold. First, the
fluid dynamics aspect of imbibition is itself very complex, and should
not be ignored in any discussion of the statistical fluctuations of the
interface. We feel that this has not been properly done so far.
Secondly, we present in a companion paper the results of a line
of investigation of spontaneous imbibition based on a phase field
formalism \cite{Dube_2005}; the present paper is intented to lay down 
the aspects which we believe essentials to any models 
of imbibition.

We start in Section \ref{macroscopicfeatures} by
reviewing the macroscopic properties of imbibition and show that even
the simple propagation of an imbibition front may have several
dynamical regimes, depending on the design of the experiment. We also
show that the macroscopic progression of the interface will have a very
strong influence on the roughening process. Section \ref{previousmodels} 
outlines the models that have been proposed for imbibition experiments as 
well as their predictions. In Section \ref{experimentalefforts}, we discuss 
the existing experimental data, and point along the way to several 
features missed by these experiments. Finally we conclude with a 
discussion and proposals concerning future experiments.

\section{Macroscopic Features of Imbibition}
\label{macroscopicfeatures}

Although extremely familiar to researchers working in the field of flow
in porous media, the details of imbibition would seem to be relatively 
unknown to the statistical physics community. It is generally defined
in reference to two-fluid flow in porous media, and corresponds to the
displacement of the lesser wetting fluid by the more wetting one 
\cite{Scheidegger_57,Sahimi_93}. Notice that this definition is
irrespective of whether the flow of the fluids is spontaneous or
induced (eg., by a pump). In this work, we will restrict ourselves to
the case of {\it spontaneous} imbibition. The flow of the fluids is
thus driven solely by capillary forces, with gravity and/or
evaporation being the only external influences on the fluids' motion.

\subsection{Capillary Rise}
\label{capillaryrise}

The simplest example of spontaneous imbibition is capillary rise: a
part of a capillary tube, of radius $R$, is immersed into a reservoir,
exposed to an ambient  
atmospheric pressure $P_0$. We assume that the fluid wets the capillary 
so that a meniscus, described by the surface $z-h(x,y,t)=0$ is formed.
At equilibrium, this surface is characterised by a contact angle 
$\theta$, obtained from Young's law \cite{Rowlinson_82}
\begin{equation}
\gamma_{lg} \cos \theta = \gamma_{sg} - \gamma_{sl},
\end{equation}
where $\gamma_{ij}$ is the surface tension between the phases $i$ and
$j$, and $s$, $l$ and $g$ stand for solid, liquid and gas
respectively. If the meniscus is in motion, the contact angle differs
from its equilibrium value \cite{deGennes_85}, but we neglect this
effect here. The important point is that the curvature causes a
pressure difference $\Delta P \equiv P_c = P_0 - P(z=h^-) = 2
\gamma_{lg} \cos \theta /R$ where $\gamma$ is the surface tension of
the liquid-gas interface and $R/ \cos \theta$ is the curvature of the
meniscus. We refer to $P_c$ as the capillary pressure. 

The motion of the fluid, of density $\rho$ and viscosity $\eta$, 
can be treated within the assumption of Poiseuille flow
\cite{Landau_fl}, i.e.\ the full Navier-Stokes equation is replaced by
the simpler Stokes equation for an incompressible fluid under the
influence of the gravitational force $\rho {\bf g}$. As usual, the
pressure field is found from Laplace's equation $\nabla^2 P = 0$, but
with boundary condition $P(z \! = \! 0) = P_0$ and $P(z \! = \! h) =
P_0 - P_c$. (We can neglect corrections to the pressure field close to
the meniscus, if its height is a lot larger than the radius of the
tube). This yields $ P(x,y,z) = P(z) = P_0 - P_c z/h(t)$, and it is
straightforward to obtain the progression of the interface,
\begin{equation}
\frac{ d h}{d t} = \frac{\kappa}{\eta} \rho g 
\left( \frac{h_{eq}}{h(t)} - 1 \right)
\label{washburn}
\end{equation}
a classical result derived by Washburn \cite{Washburn_21} and Rideal
\cite{Rideal_22}. Washburn's and Rideal's equation includes the
permeability of the tube $\kappa = R^2 / 8$ and the equilibrium height
of the meniscus $h_{eq} = P_c / \rho g$. Studies of capillary rise
including the inertial term of the Navier-Stokes equation
\cite{Szekely_71} show that Eq.\ (\ref{washburn}) is essentially
correct, with the exception of very short times. Notice also that this
equation of motion also neglects completely the problem of the actual
motion of the contact line between the gas-solid and liquid phases
\cite{deGennes_85}.

The (transcendental) solution of Eq.\ (\ref{washburn}) has the
following asymptotic properties: Defining $h(t_0)$ as the initial
height of the column and $\tau_{eq} \equiv h_{eq} \eta / \kappa (\rho
g)^2$ as an equilibration time, for low heights $h \ll h_{eq}$ (and also
in the absence of gravity where $h_{eq} = \infty$) the rise is of the
form
\begin{equation}
h^2 (t) - h^2 (t_0) \; = \; \frac{\kappa P_c}{\eta} \; (t - t_0).
\label{t-half}
\end{equation}
The equilibrium height is approached by $h$ exponentially
\begin{equation}
h (t) \; \sim \; h_{eq} ( 1- e^{-t/\tau_{eq}}).
\label{rise-eq}
\end{equation}

This was examined experimentally by Washburn \cite{Washburn_21} and
Rideal \cite{Rideal_22} as well as numerous others, who found very
good agreement between theory and experiment. Capillary rise in a
liquid--liquid system was also examined by Mumley et al.\
\cite{Mumley_86}. They confirmed Washburn's result in the case of
perfect wetting, but reported discrepancies for systems with non-zero
contact angle. In this case, the rise was {\it slower} that $t^{1/2}$,
a behaviour attributed to the motion of the contact line
itself \cite{deGennes_85}. 

\subsection{Capillary Rise in Porous Media}

With the basic capillary rise phenomenon understood, we can now turn
to spontaneous imbibition in porous media. The flow of liquids in
porous media is generally described in terms of Darcy's equation
\cite{Scheidegger_57,Sahimi_93} 
\begin{equation}
\langle {\bf Q} \rangle = - \rho A \frac{\kappa}{\eta} 
( {\mbox {\boldmath $\nabla$} } P - \rho {\bf g} ),
\label{darcy}
\end{equation}
where $ \langle {\bf Q} \rangle$ is the average mass of fluid
transported per unit time through the cross-section $A$ of the porous
medium, and $\kappa$ and $\eta$ are the average permeability and
viscosity respectively. Darcy's equation arises from an averaging
procedure of the porous medium and ignores all details on length
scales smaller than the pores. If we assume the porous medium to be
homogeneous, such that the permeability is a constant $\kappa_0$
independent of the fluid concentration, we can still solve Laplace's
equation for the pressure and obtain a coarse-grained pressure
gradient, ${\mbox {\boldmath $\nabla$} } P = {\bf \hat{z}} P_c / h(t)$,
with $P_c$ an effective capillary pressure, and $h$ the average height
of the fluid column. This is of course valid only when the notion of
an interface is itself well defined, i.e.\ it should not be too
``fuzzy'' up to macroscopic scales. Under these assumptions, and with
the identification $\langle Q \rangle = \rho A \,dh /dt$, Darcy's
equation leads directly to the Washburn--Rideal result, Eq.\
(\ref{washburn}) with a dynamical rise given by Eqs.\ (\ref{t-half})
and (\ref{rise-eq}).

This form of Darcy's equation has been used to study many experiments 
of fluid propagation in porous media, including some fibrous materials
(see below). The agreement is not always perfect -- we shall come back
to this in Section \ref{validity}. At this point, a more complete system of 
equations \cite{Sahimi_93,Hilfer_98} would only obscure the physics.
 
\subsection{Evaporation}

Considering the typical experimental setups for imbibition front propagation
\cite{Amaral_94}, evaporation effects should also be included to
Washburn's equation. As far as we are aware, no detailed studies of
fluid motion through thin porous media including the effects of evaporation
have been performed. A natural assumption is to introduce an evaporation 
rate proportional to the area of the fluid exposed to air, i.e.\ the net
loss of fluid mass per unit time due to evaporation is $Q_e = - 2
{\cal E} \rho L h (t)$ where $L$ is the lateral width, and ${\cal E}$
is a phenomenological evaporation rate  per unit area. Thus Eq.\
(\ref{washburn}) is modified to 
\begin{equation}
\frac{ d h}{d t} = \frac{\kappa}{\eta} \left( 
\frac{p_c}{h} - \rho g \right) - \epsilon h
\label{wash-grav-evap}
\end{equation}
where $\epsilon = 2 {\cal E} \rho$ is the evaporation rate. This form
obviously neglects changes in the concentration profile towards the
surface, as well as the evaporation at the interface itself, which is
both justified for thin media such as a sheet of paper.

An immediate consequence of Eq. (\ref{wash-grav-evap}) is an equilibrium
height depending on the evaporation rate
\begin{equation}
h_{eq} = \frac{h_e^2}{2 h_g} \left( \left( 1 + \frac{4 h_g^2}{h_e^2}
\right)^{1/2} - 1 \right)
\end{equation}
where $h_g = P_c /\rho g$ and $h_e = ( P_c \kappa / \epsilon \eta
)^{1/2}$ are respectively the equilibrium heights in the cases where
only gravity or evaporation have a significant influence, with a 
crossover defined by $2 h_g \sim h_e$. When gravity can be neglected,
the height $h(t)$ behaves as
\begin{equation}
h^2 (t) = h^2 (t_0) \; e^{-2 \epsilon (t - t_0)}
+ h_e^2 \; \left( 1 - e^{-2 \epsilon (t - t_0)} \right)
\label{wash-evap}
\end{equation}
Well below $h_e$ the rise still follows $h(t) \sim t^{1/2}$, 
and it again approaches exponentially the equilibrium height. 

\subsection{Addition of a Solution}

Many experiments on the roughness of the interface in an imbibition
context have studied the behaviour of a dye, added to the pure fluid
\cite{Buldyrev_92,Amaral_94,Kwon_96}. In the literature on porous
media this is referred to as the hydrodynamic dispersion phenomenon
\cite{Sahimi_93}. At the simplest level, it can be treated by the
introduction of a dye concentration field $K({\bf x},t)$ advected by
the macroscopic velocity field ${\bf v}$,
\begin{equation} 
\frac{\partial K}{\partial t} + {\bf v} \cdot 
{\mbox {\boldmath $\nabla$} } K = D \nabla^2 K,
\label{dye_dynamics}
\end{equation}
where $D$ is a diffusion constant (in some cases, it may be necessary
to introduce a diffusion tensor $D_{ij}$ \cite{Sahimi_93}).

Two remarks on Eq.\ (\ref{dye_dynamics}) are in order. First, in
everyday life, it is easily seen that the fluid front is always faster
than the dye front. This clogging phenomenon, which may be due to a
smaller permeability for the dye, may easily be incorporated
phenomenologically by introducing a constant $\lambda < 1$ in the
front of the convective term, i.e.\ 
\begin{equation}
 {\bf v} \cdot {\mbox {\boldmath $\nabla$} } K 
\rightarrow \lambda  {\bf v} \cdot {\mbox {\boldmath $\nabla$} } K.
\end{equation}
The second point concerns the stopped interface. If the pinning of the
interface, at a distance $h_e$ from the reservoir, is due to
evaporation, there will nevertheless be a fluid motion, with
approximate velocity $ v_e \sim \kappa P_c / \eta h_e $, in order to
compensate the losses. This implies that more and more dye particles
will be brought to the fluid interface, thus creating a band of high
dye concentration, analogous to the coffee rings examined recently 
\cite{Deegan_97}. The increase of the width of this band with time 
can be used as an alternative way of measuring the relevant 
macroscopic parameters.

\subsection{Validity of Macroscopic Description}
\label{validity}

We now discuss the validity of the macroscopic results for the
specific case of spontaneous imbibition in paper. Ordinary paper is
made out of wood fibers and, in many cases, chemical additives and filler
materials like talc and clay, arranged in a disordered structure. Not
only is there a wide distribution of pore sizes, with a high effective
tortuosity but the surface of the network is extremely uneven. This
gives rise to all the standard problems in defining a static/dynamical
contact angle for inhomogeneous substrates.

Darcy's equation, Eq.\ (\ref{darcy}), is relatively well established
(to the point that it is even often referred to as a law) for general
porous media. However, paper, as well as other fibrous materials
present the  peculiarity that the fiber structure may be modified by
the contact with the liquid, a phenomenon known as swelling
\cite{Bristow_71}. Cellulose fibers show a great affinity to water and
can absorb  large quantities in millisecond timescales, giving rise to
concomitant changes in fiber volume and pore structure (this is
however not the case for many organic fluids and oils). Thus one's
intuitive picture of fibers as capillary tubes is false: the pore
structure is highly non-trivial and in some cases time--dependent.
There are two different effects that play a role: the volume to be
filled with liquid increases, and the flow resistance of the pore
network changes.

There can also be an exchange of liquid between the inside of a fiber and
the ``surface'' pores, which complicates enormously the fluid flow, since
there are no well defined ``structures'' (either the pores, or the
fibers) responsible for the capillary forces \cite{Aspler_87,Pezron_95}. 
This is of course a very serious impediment to any kind of attempt at a
``microscopic description'' of the structure, say in terms of a
percolation network \cite{Lenormand_88}. 

A more serious problem with water is the fact that it is not actually
known whether a pressure balance argument is always valid. On short
time scales (smaller than a few seconds), it has been proposed that
front penetration would proceed in pores first with the establishment
of a prewetting layer through {\it diffusion} \cite{Salminen_88}, in
which case the front position could advance as $t^k$ with $k$
varying up to unity. Experiments have demonstrated that clear,
Washburn-like behavior can be obtained in conditions that amount to
basically no interaction (swelling, prewetting) between the material
and the penetrating liquid \cite{Gillespie_58}. Note that minute
applications of chemicals, e.g.\ during paper manufacture, may induce drastic
changes in the effective viscosity or surface tension of the invading
liquid.

There are however some advantages to paper. The very high permeability
of the fibers will reduce the formation of overhangs in the interface.
Darcy's equation cannot treat trapped air bubbles in bulk porous media
\cite{Hilfer_98}, which in paper only play a minor role: paper is thin and 
the pores should be connected well to the surface. Moreover
spontaneous imbibition is slow and allows more time for the removal of
overhangs and trapped bubbles.  

In conclusion, one would believe that a well defined fluid-air
interface should exist, and that Darcy's equation should be valid on
length scales larger than the interface width, provided that
interactions between the liquid and the fibers are minimal. In that
sense, a full hydrodynamical treatment of the problem may not 
be necessary (as noted in \cite{Dube_2005}). In many
cases, it may also be necessary to introduce a concentration dependent
permeability, i.e.\ $\kappa = \kappa (\rho)$ \cite{Gillespie_58}. In
any case a Washburn approach may be a first step. In that respect,
recent experimental work with paper and {\it deionised} water, 
showing front propagation consistent with Washburn's equation is 
quite encouraging \cite{Zik_98}. 

\subsection{Statistical Fluctuations of the Interface}

So far, we have considered a flat interface in a completely homogeneous
medium. In general, the disorder in the paper causes fluctuations
which will accumulate to roughen the advancing front. In the standard
picture of kinetic roughening \cite{ref_gen,Krug_97} 
the fluctuations of the interface are
correlated up to a distance $\xi_{\|} (t)$, a lateral correlation
length, increasing in time as $\xi_{\|} \sim t^{1/z}$ and described
by the dynamical exponent $z$. The vertical extent of the interface
fluctuations, its width $W$, is related to $\xi_\|$ through the
roughness exponent $W \sim \xi_\|^\chi$. Thus the width increases with
time as $W \sim t^\beta$ with $\beta = \chi / z$, until the interface
fluctuations have saturated, ie.,  until $\xi_\| (t) = L$, 
which defines the crossover time $t_\times \sim L^z$.

The initial increase of the width and eventual saturation are comprised in
a (Family--Vicsek) scaling form
\begin{equation}
W (L,t) = L^{\chi} f (t/L^z)
\end{equation}
with the scaling function $f(u) \sim u^{\beta}$ for $u \ll 1$ and
$f(u) \sim const$ for $u \gg 1$. 

The spatial height difference correlation function (of the $q$th moment)
\begin{equation}
G_q ({\bf r},t) = \langle | h({\bf r}+{\bf r}',t)-h({\bf r}',t) |^q
\rangle^{1/q}
\end{equation}
generally scales similarly to the total width, i.e.\ $G_q \sim r^\chi$
for small $r$, and approaching a constant as $r \to
\infty$. However, care has to be taken: In the case of {\em anomalous}
scaling the height differences for fixed $|{\bf r}|$ do not saturate
with time, and $G_q$ shows the {\em local} roughness
exponent $\chi_{loc}$ \cite{Krug_97,Lopez_97}. The height
difference distribution for fixed $|{\bf r}|$ may also have a long tail,
causing intermittent or ``turbulent'' jumps in the height
configuration in which case each moment has a different exponent
$\chi_q$, ie., the interface exhibits multiscaling \cite{Krug_94}.
Also a temporal height difference correlation function
\begin{equation}
C_q (t) = \langle | h({\bf r},t+t')-h({\bf r},t')-(\bar
h(t+t')-\bar h(t')) |^q \rangle^{1/q}_{{\bf r}}
\end{equation}
can be considered (e.g.\ in \cite{Horvath_95}), which is of particular
use in the case the system exhibits time--translational invariance.

To our knowledge none of the works studying interface fluctuations
in imbibition 
have checked for anomalous (as determined by the higher
moments of the correlation functions) types of scaling behaviour
\cite{Buldyrev_92,Barabasi_92,Family_92,Buldyrev_ph,Amaral_94,Horvath_95,Kumar_96,Kwon_96,Zik_97},
although this is in principle difficult, since an exact diagnosis 
for higher moments $q>2$ is severely hampered by large statistical
fluctuations. These concepts are nevertheless important  when 
the randomness of the medium is described with random-field disorder
\cite{Dube_2005}.

Another peculiarity of imbibition is that, apart from 
the standard kinetic lateral correlation length $\xi_\|(t)$,
arising from the accumulating history of fluctuations, another lateral 
length scale is to be seen: local conservation of the fluid
determines a lateral length $\xi_\times$ related to the average motion of
the interface. Intuitively, it is clear that such a length scale must exist,
since fluctuations ahead and behind the average position of the
imbibition front have respectively a slower and faster instantaneous
velocity. To make a quantitative argument, we introduce an effective
surface tension $\gamma^*$, representing the energy cost of a curved
air-liquid interface on a {\it macroscopic} scale \cite{Krug_91}. Any
curvature at the interface -- the typical size of fluctuations being
$W$ vertically and $\xi_\times$ laterally -- modifies the pressure by an
amount $\Delta P = \gamma^* W / \xi_\times^2$ (the Laplace pressure
effect). This should be compared with
the difference in the pressure field across the same vertical distance
$W$ due to the pressure gradient derived in Section
\ref{capillaryrise}, namely  $\Delta P = P_c W / H$, from which we obtain
\begin{equation}
\xi_\times^2 \sim \gamma^* H / P_c,
\end{equation}
relating the parallel length scale $\xi_\times$ to the height of the
interface. Beyond this length scale we do not expect any correlated
roughness of the interface, since those fluctuations would be
suppressed by the overall gradient in the pressure field $P_c/ H$.
 
Let us recall once again that the interface continuously slows down
\cite{note_4}, without gravity nor evaporation $H \sim t^{1/2}$. Thus
the slow increase of $\xi_\times \sim H^{1/2} \sim t^{1/4}$ with time
leads to an increase of the width $W \sim t^\beta$ with $\beta =
\chi/4$. However, we stress that this increase is conceptually
different from the increase of the lateral correlation length with a
dynamic exponent $z$ in standard models of kinetic roughening.

\section{Models of Imbibition}
\label{previousmodels}

In parallel with experimental work, many theoretical models of imbibition
have been developed. These models fall either in  discrete (cellular
automata) or continuous classes. Most of the theoretical discussion on
imbibition was done in terms of the DPD model, which was shown to
belong to the same universality class as the quenched KPZ equation.
Other types of models or continuum equations have also been proposed.

\subsection{Cellular Automata Models}

The earliest theoretical investigations of imbibition phenomena were
based on the {\em Directed Percolation Depinning} (DPD) model
\cite{Buldyrev_92,Barabasi_92,Amaral_94}. It is a cellular automaton,
with space discretized in cells which are either blocked or wettable,
with the blocked cells being a fraction $p$. The particularity of the
model is that overhangs in the interface are removed as soon as they
occur, which thereby introduces anisotropy in the interface. Growth
takes place by invasion of available cells until the interface comes
across a percolating directed path of blocked cells. Clearly, the
statistical properties of the interface are related to the
properties of the percolating path. Close to the percolation threshold
$p_c \simeq 0.47$, the directed percolating path is characterized by the
parallel and perpendicular length scales $\xi_{\|} \sim | p- p_c|
^{-\nu_{\|}}$ and $\xi_{\bot} \sim | p- p_c| ^{-\nu_{\bot}}$, with
values $\nu_{\|} \simeq 1.7$ and $\nu_{\bot} \simeq 1.09$. so the
width $ W(L) \sim \xi_{\bot} \sim L^{\nu_{\bot}/\nu_{\|}}$, yielding
$\chi \simeq 0.63$.

To treat evaporation, the DPD model was modified by the introduction
of height dependent fraction of blocked cells, i.e.\ $p = p(h) \sim 
h /h_0$ with an equilibrium height $h_0 \sim (\nabla p)^{-1}$ 
defined by $p(h_0)=p_c$ \cite{Amaral_94}.
In this model, the interface is described by the standard DPD roughness
exponent up to a length $l_{\times}  \sim (\nabla
p)^{-\gamma/\chi_{DPD}}$ after which it saturates to a constant value
$w_{sat} \sim ( \nabla p)^{-\gamma}$. The
identification $w_{sat} \sim \xi_{\bot}$ then yields $\gamma =
\nu_{\bot} / (1+\nu_{\bot}) \sim 0.52$, a value comparable to the
experimental results. 

As already noted \cite{Horvath_95}, there is a physical justification
to the main feature of the DPD model, namely the erosion of
overhangs. It is plausible that propagation of the fluid in a direction
parallel to the reservoir, if it has a chance to occur, will be
favoured over the motion of the fluid away from it, since this does
not result in any change in the average capillary pressure. In that
sense, any overhangs will eventually be removed, although the question
of the different time   scales involved in the dynamics is certainly
extremely complex \cite{Dube_2005}.  

The DPD model in its simplest form then probably describes well the
statistical properties of the pinned interface. It is however
difficult to believe that this model can describe the whole 
dynamical motion of the interface, since it does not account for
liquid conservation. It already fails by predicting a {\em constant}
average interface velocity. As we saw in section
\ref{macroscopicfeatures}, one of the principal features of imbibition
is a constant slowing down of the front, a result intimately related
to fluid conservation. It can be argued that the DPD refers
specifically to the motion of the dye, but the rise of the dye front
is necessarily bounded by the liquid-air interface, and the
conservation law governing the motion of the fluid must be reflected
on the motion of the dye particles. From that point of view, simply
replacing ${\bf v} \sim t^{-1/2}$ in Eq. (\ref{dye_dynamics}) certainly
would be incorrect.

It is interesting to show how the global behaviour of the
modified DPD model may be related to the macroscopic Washburn behaviour,
Eq. (\ref{wash-grav-evap}).
The density of {\it open} cells $1-p(h)$ can be thought of as an
effective force acting on the interface, similar to the gradient in
pressure on fluid motion.
Since the Washburn Eq. is by essence a dissipative equation, we can
associate, {\it away from pinning}
\begin{equation}
 p (h)  \rightarrow
\frac{\kappa}{\eta} \left(
\frac{p_c}{h} - \rho g \right) - \epsilon h
\end{equation}
The assumption of constant gradient thus effectively corresponds to an
evaporation rate. It is also possible, at the macroscopic level, to
include gravity, and to mimic the slowing down of
the interface. Note however that we predict a pinning height
$h_e \sim \epsilon^{-1/2}$ while the assumption of Amaral et al.
implies $h_e \sim (\nabla p)^{-1}$, since they neglect the capillary
driving term.

Another model of evaporation was introduced by Kumar and Jana 
\cite{Kumar_96}. The model is essentially similar to the DPD model,
with the modification that each cell is not necessarily ``full'' or
``empty'', but may contain many ``subunits''. Evaporation is modeled
by a loss of $n$ subunits in the transfer between cells. It is found
that below a critical $n_{c1}$, the interface propagates
indefinitely. Between this value and a second critical loss rate
$n_{c2}$, they observe an interface behaviour similar to the one
observed by Amaral et al.\ \cite{Amaral_94}, but with value $\chi =
0.5$ and $\gamma = 3.0$. For $n > n_{c2}$, this scaling regime breaks
down and the roughness exponent becomes $n$--dependent. These results
were backed experimentally, but not in any consistent way.

It is rather difficult to believe that indefinite front propagation
under evaporation (obtained for $n < n_{c1}$) is physically
realistic, nor seems the way of including evaporation convincing. In
this model, the concentration of fluid molecules decreases 
continuously from a maximum $N_0$ to a value of $0$ at the interface,
which seems physically unrealistic.

We terminate by presenting invasion models based on the random field
Ising model \cite{Martys_91}, which predict a depinning transition, as
well as a change of morphology of the interface at some length scale
related to the capillary pressure. It seems however unsuited for
spontaneous imbibition, because any advance of the front requires the
increase of an externally applied pressure.

A variation of this idea, introduced by Sneppen \cite{Sneppen_92}, 
allows invasion always at sites of lowest resistance. An interesting
aspect of this model is temporal multiscaling \cite{SnepJen_93}, due
to avalanche motion \cite{Olami_94,Leschhorn_94}. Avalanches are 
also present in the process of spontaneous imbibition, although
the consevation law imposes a natural cutoff on their size and 
distribution \cite{Dougherty_98,Dube_2005}. It is nevertheless 
interesting that a similar lack of temporal scaling is also seen
in the phase field model of imbibition, thus surviving the 
introduction of a conservation law.

\subsection{Continuum Description}

The DPD model (in its original version without ``evaporation'')
belongs to the same universality class as the quenched KPZ equation
(see e.g.\ Chapter~10 in \cite{ref_gen} and references therein, as well
as \cite{Csahok_93}) 
\begin{equation}
\frac{\partial h(x,t)}{\partial t} = \nu \nabla^2 h(x,t) -
\frac{\lambda}{2} (\nabla h(x,t))^2 + \eta(h(x,t),x).
\end{equation}
A related equation with {\it multiplicative} noise has been introduced
by Csah\'ok et al.\ \cite{Csahok_93} in order to model the random
porosity of the medium. Since these equations are local at the
interface, they do not contain a conservation law for the fluid.

%
%

With these considerations in mind, a continuum equation was introduced
by Zik et al.\ \cite{Zik_97} following Krug and Meakin
\cite{Krug_91}. In Fourier space, it has the form
\begin{equation}
\frac{dh_k}{dt} = -\frac{\kappa p_c}{\eta} |k| \frac{h_k}{h_0(t)}
+ \eta_k (t)
\label{cont-eq}
\end{equation}
where $k$ is the wave-vector, $h_0 (t)$ is the mean height of the
interface at time $t$ (cf., Eq.\ (\ref{t-half})), and $\eta_k$ is the
noise term, assumed to be annealed, with correlations $\langle
\eta_k(t) \eta_{k'}(t') \rangle \sim \delta_{k+k'} \delta(t \! - \!
t')$.  This equation is essentially similar to the one derived by Krug
and Meakin for the roughness of stable Laplacian fronts with noise
and corresponds to the leading term of a Saffmann-Taylor
analysis of the problem. Note that $|k|$ is nonlocal in space.

The other difficulty with respect to spontaneous imbibition, the
quenched nature of the disorder, remains. As long as the interface
sweeps fast enough through the medium, the disorder acts as annealed,
time dependent noise \cite{Nattermann_94,Leschhorn_97}, although the
question of the effective noise correlator is far from trivial 
\cite{Dube_2005} --- Eq.\ \ref{cont-eq}, as it stands, gives the unphysical
results of an interface that {\it never} saturates. In any cases, at late 
times under slow average interface propagation in a continuum model one is 
confronted with both the difficulties of a {\em nonlocal} interface 
equation with {\em quenched} disorder.

\section{Experimental Efforts in Imbibition}
\label{experimentalefforts}

Most of the experiments concerned with front propagation in random
media seem to produce a self-affine interface, but the numerical value
of the exponents often bears little resemblance to the standard
universality classes and/or associated models. Imbibition of porous
media would seem like an obvious first experimental choice, since
the associated time and spatial scales are easily accessible in the
laboratory. Many such imbibition experiments were performed, but it
turns out that there is very little in common between the different
experiments. This is partly due to the fact that different experiments
had different goals, but it is also a reflection of the great complexity
of the processes involved in imbibition.

\subsection{Experiments on Pinned Interfaces}

The earliest statistical physics imbibition experiments were done by Buldyrev et al.\ 
\cite{Buldyrev_92,Barabasi_92,Buldyrev_ph} and Family et al.\
\cite{Family_92}. The first experiment was performed with a dye
solution in a vertical capillary rise setup. The rising front moved
from the reservoir and the roughness of the interface developed during
the process. Eventually, the dye front stopped, due to gravity and/or
evaporation (no dynamical measurements, either micro- or macroscopic,
were done in this experiment), and the roughness of the pinned
interface was measured. The main experimental conclusion of the paper
was a roughness exponent with value $\chi = 0.63$, consistent with the
DPD model (see above, Section \ref{previousmodels}). However, the
length scale over which the scaling behaviour was observed is
extremely small. For a sheet of paper of total lateral extent $L = 40$
cm, $C(l) \sim l^{\chi}$ only for length scales $l$ smaller than
$l_{max} \sim 1$ cm, after which it levels off to a constant
or to a logarithmic function of $l$ \cite{note_1}. Since this scaling
region is only a few times larger that the length scale of of the
fibers themselves, it is not at all clear whether the exponent $\chi$
really is a universal value, or simply results from the microscopic
fiber structure. As a  comparison, the scaling region for the paper
burning experiment \cite{jyvkl_97} was from $ 1.5$ cm to about $10$
cm, for a total length $30$ cm \cite{note_2}. A similar experiment was
done in a three dimensional sponge-like material
\cite{Buldyrev_ph}. Again, the stopped interface was observed,
yielding a roughness exponent $\chi^{(2d)} \sim 0.5$, a result 
consistent with the higher-dimensional version of the DPD model. No
other experimental details were however given. 

In a further set of experiments, Amaral et al.\ \cite{Amaral_94}
studied the role of evaporation in more detail. They again considered
the stopped interface, but changed the height where pinning occured
through the evaporation rate (presumably by modifying the humidity
during the experiment, although this is not specified). Their main
result is that the width of the pinned interface is related to the
pinning height $H_p$ (and thus to the evaporation strength) through a new
exponent $\gamma$, namely $w_{sat} \sim H_p^\gamma$. The experiments
gave a value $\gamma = 0.49$, which -- as reported in Section
\ref{previousmodels} -- can also be obtained from a modified version
of the DPD model \cite{Amaral_94}. (We will come back to this in
Section \ref{Analysis}). On the other hand, Kumar and Juma
\cite{Kumar_96} also performed an experiment in presence of various
evaporation rates and claimed that the roughness exponent was not
universal but stronly dependent on the evaporation rate, i.e.\  $\chi
= \chi (\epsilon)$, although no systematic experimental investigation
of this breakdown of universality was done.

\subsection{Experiments on Moving Interfaces}

The experiment of Family et al.\ was performed in an horizontal
capillary setup with water only, and the position of the air--water
interface was recorded both temporally and spatially
\cite{Family_92}. The main experimental results were an average
interface progression $\bar{h} \sim t^{\delta}$,  with $\delta = 0.7$,
and a self--affine interface described by a Family-Vicsek scaling
relation, and characterized by the exponents $\beta = 0.38$ and $\chi
= 0.76$. The first result $h(t) \sim  t^{0.7}$ is certainly
intriguing. However, it should be noticed that a reservoir was placed
underneath the piece of paper, in order to prevent evaporation. It is
of course possible that this caused condensation instead, thus
increasing the velocity of the front. It is also in line with the
results of non--Washburn water penetration in paper
\cite{Salminen_88}. Also here the spatial scaling regime was rather
small, for distances below $l_{max} \sim 2$ cm for a $40$ cm wide
sheet of paper. It should also be noticed that these experiments were
made with chinese paper, having a thickness of only a few fiber layers, 
which may also influence the results. 

The temporal scaling of the interface was studied in details by
Horv\'ath and Stanley \cite{Horvath_95}. Moving a piece of paper such
that the interface always remained at a fixed distance $H$ above a
reservoir, they found a power law behaviour for the time correlation
function $C_2(t) \sim t^{\beta}$ with $\beta = 0.56$. Another result
is that the velocity $V$ at which the the piece of paper must be moved
towards the reservoir in order to keep the interface at a constant $H$
varies as $V \sim H^{-\Omega}$, where $\Omega = 1.6$. Notice how this
implies an interface propagating such that $H(t) \sim t^{2/5}$, since
$V = dH/dt$. This is slower than $\delta = 1/2$ expected from Darcy's
equation, but consistent with our earlier discussion. Unfortunately,
the scaling behaviour of the interface as a function of space was not
discussed and/or measured in this otherwise very careful work.

It is certainly interesting that the value of $\beta$ is constant for
all heights considered. The different heights did however affect the
time at which the time correlation functions saturated, and the values
at which it did so. Horv\'ath and Stanley found for $C(t)$ a scaling
form
\begin{equation}
C(t) \sim V^{-\Theta_L} f (t V^{(\Theta_t+\Theta_L)/\beta} ),
\end{equation}
where $f(y)$ is a scaling function such that $f(y) \sim y^{\beta}$ for
$y \ll 1$ and $f(y) \sim const$ for $y \gg 1$. The values of the
independent exponents were $\Theta_L=0.48$ and $\Theta_T=0.37$
\cite{note_3}.

Another experiment was performed by Kwon et al.\ \cite{Kwon_96}. They
deposited a paper towel on an inclined glass plate and followed the
capillary rise of a dye solution, giving a mean interface dynamical
propagation  $\bar{h} (t) \sim t^{0.37}$. Again, two scaling regimes
were present in the saturated width; on small length scales ($\leq 2$
cm), $\chi = 0.67$ while on larger length scales (up to $20$ cm) $\chi
\sim 0.2$. Within the simple Family-Vicsek picture they obtained
$\beta = 0.24$ on the short lengthscale regime.

Finally, Zik et al.\ \cite{Zik_97} performed an horizontal capillary
front experiment. They obtained rough interfaces only with highly
anisotropic paper, for which they found $\chi = 0.4$. For isotropic
paper, the roughness was at best logarithmic. It is remarkable, that
the scaling for the anisotropic paper was observed through a large
range of length scales, not only up to fiber length.

\subsection{Analysis of the Experiments : Influence of Fluid
Conservation}
\label{Analysis}

As the previous paragraphs show, there is a large amount of
experimental work on the subject of spontaneous imbibition and front
roughening, but a coherent picture does not emerge naturally. The
results are quite contradictory and difficult to analyse properly. 

We have already mentioned a few experimental difficulties inherent to
imbibition in paper, the most important being the complicated flow
profile of a fluid inside the paper, due to fiber swelling and
dissolution. Certainly all experiments suffer from these difficulties
at least to some degree. Apart from the work of Family et al.\
\cite{Family_92}, all experiments are consistent with an average front
propagating roughly according to the Washburn--Rideal result, Eq.\
(\ref{washburn}). They don't agree perfectly, but definitely reflect
the influence of the conservation of fluid on the propagation process.

One consistent feature to emerge from the experiments is an interface
which has developed roughness only on very short length scales. A
partial remedy to this problem is, as done by Amaral et al.\
\cite{Amaral_94}, to introduce a length scale $l_p$ above which
the interface saturates. It their experiment and model (see also
Section \ref{previousmodels}) it becomes visible in the width the
interface takes at the height where it gets pinned by evaporation, and
they establish the relation $w \sim H_p^\gamma \sim 
l_p^\chi$. 

The idea of such a length scale has of course a wider range of 
applicability than in the context of imbibition with evaporation. 
Indeed, the experiment of Horv\'ath and Stanley \cite{Horvath_95}, 
done with little influence of evaporation, and no discernable influence 
of gravity, showed the existence of a similar length scale, in this 
case, related to the velocity $V$ at which paper was being pulled 
down (or, equivalently, to the average height of the interface).

We have already shown that such a length scale $\xi_\times (H)$, 
connected to the existence of a conservation law existed. 
It can provide a natural 
connection between the work of Amaral et al.\ and Horv\'ath and
Stanley, provided that one is willing to revise the role of
evaporation. If the sole role of evaporation is to stop the
macroscopic progression of the interface, with little or no influence
on the statistical fluctuations of the interface itself, then the
width  $w(t) \sim \xi_\times^{\chi} (H_p) \sim H_p^{\chi/2}$ and therefore
$\gamma = \chi/2$. 
This length should also influence the setup examined by Horv\'ath 
and Stanley \cite{Dube_2005}. 
Indeed, it provides an alternative explanation to the
velocity (or alternativley height) dependence of the saturated value
of the time correlation function, since
\begin{equation}
C^2_2 (\tau \rightarrow \infty) \sim \int_{1/\xi_\times (H)}^{1/a} 
\frac{dk}{k^{2\chi+1}} 
\end{equation}
which yields $C_2(\tau \rightarrow \infty) \sim H^{\chi/2} \sim 
V^{-\chi/2}$. 
Note that as long as $\xi_\times < L$, the
correlation function $C_2(t)$ is {\em independent} of the total
length of the system. Since only one length of paper was considered
experimentally, it is unfortunately impossible to check this hypothesis.

However, in order to demonstrate convincingly the existence of such a
length scale (or of any other length scales) and the scaling below it, 
it is imperative to extend the region of scaling. 
We fail to see how a sub-centimeter description
of the roughness might be considered universal. It is quite possible
that the crossover $\xi_\times$ associated with paper really is in
the sub--centimeter range, in which case, the alternative way to check
for scaling is in the scaling of the width. Even in the freely
rising case, the total width of the interface should show some early
time power law behaviour, with some exponent $\beta$.

One should also note the fact the structure of paper does have
short-scale power-law correlations \cite{Pro96}. These are
manifest in the two-point-correlation function of the areal
mass density of the paper, with the cut-off of the effective
power-law correlations extending in practice to a few times the fiber
length. It is by no means clear what role such correlations play in
imbibition, since penetration of the liquid may take place both
inside the sheet, and on the outer surface. Also, it is not clear how
these correlations map into correlations in the hydraulic conductivity
or local permeability. They however remind of the fact that scaling
exponents established on lengthscales that are also relevant from the
microstructural viewpoint should be taken with a grain of salt. 
Still, it is quite possible that the effective noise
correlator at the interface will display power law behaviour, as in
the case of forced flow through Hele-Shaw cells \cite{Horvath_91}.

\section{Summary and Perspective}

In this paper, we have tried to extract a coherent picture of all
previous studies about interface roughening in spontaneous imbibition
experiments. Concerning theoretical understanding, the DPD picture
plausibly reproduces the properties of the pinned interface. We
strongly feel that not enough attention has been paid to the
macroscopic dynamical features and their connection to the underlying
microscopic structure. The flow of fluid in paper is influenced
by many microscopic mechanisms, and a study of the average front
velocity already indicates which ones may or may not be relevant.  
The main experimental characteristic of imbibition certainly is the
slowing down of the front, a phenomenon not seen in other
systems. Some very simple predictions concerning evaporation and
gravity would already allow a solid comparison between the expected
behaviour and the experimental results.

An important reason to consider the fluctuations of the propagating
front is the presence of a lateral length scale $\xi_\times$, for
which experimental evidence already exists
\cite{Amaral_94,Horvath_95}, and which has a very strong influence on
the roughening process.

Finally, we believe that a complete description of imbibition 
(including the dynamic interface) must include a conservation law
for the whole fluid, not simply at the interface level. This is for
two reasons, first the average interface ought to slow down
continuously, and secondly the dynamics of the interface need to have
long range interactions. Also, any model must deal the quenched
nature of the noise in a suitable way. These requirements make it 
impossible to describe the interface roughening with a {\em local}
equation of motion. To answer these criteria, we introduce in the
next paper \cite{Dube_2005} a phase--field model of imbibition,
which incorporates explicitely a conservation 
law for the liquid in a disordered medium, 
and thus produces the correct macroscopic physics.


We finish this overview by pointing out a few experimental
directions. The use of ``ordinary'' paper for the medium creates some
problems that are usually neglected. In this respect it comes to mind
that other liquids than water or water/ink mixtures might well prove
advantageous, in that they are less susceptible to evaporation and/or
do not interact with the fibers themselves. Another option is to
monitor the roughness of the front using an Hele-Shaw cell. This
eliminates the problem of fiber-liquid interaction.  

\section{Acknowledgements}
We wish to acknowledge interesting and stimulating discussions with 
J.  Kert\'esz, T. Ala-Nissila, K. R. Elder, S. Majaniemi, and J. Lohi. 
This work has been supported by the MATRA program of the 
Academy of Finland.
%
%
%


\begin{thebibliography}{99}
\bibitem{note_1} This can be seen easily from Fig.\ (2) of
                   Ref.\ \cite{Buldyrev_92}.
\bibitem{note_2} On length scales smaller than $1.5 \, \mbox{cm}$, there
                 may be signs of scaling, with a $\chi \sim 0.7$, but
                 again, it is difficult to attach any universal
                 significance to length scales of the order of the fiber
                 length.
\bibitem{note_3} Horvath ands Stanley also included the total system size 
                 $L$ in a scaling hypothesis, but since only one paper size 
                 was used, their hypothesis could not be fully tested.
\bibitem{note_4} As far as we are aware, only the experimental groups
                 of Refs. \cite{Family_92},  \cite{Kwon_96} and 
                 \cite{Zik_97} have realised this physical fact, but 
                 the implications of it were not discussed in any
                 elaborate ways.
\end{thebibliography}
\end{document}